\documentclass[a4paper,11pt]{article}
\usepackage{enumitem}
\usepackage{setspace}
\usepackage{pos}

\renewcommand{\paragraph}[1]{\noindent \textbf{#1}}
\newcommand{\codeStyle}[1]{\texttt{#1}}

\title{CRPropa 3.2: a framework for high-energy astroparticle propagation}
\ShortTitle{CRPropa 3.2}

\author*[\Radboud]{Rafael {Alves Batista}}
\author[\Bochum, \RAPP]{Julia {Becker Tjus}}
\author[\Bochum, \RAPP]{Julien {D\"orner}}
\author[\GSSI]{Andrej {Dundovic}}
\author[\Bochum, \RAPP]{Bj\"orn {Eichmann}}
\author[\Bochum, \RAPP]{Antonius {Frie}}
\author[\RWTH, \MPIFR]{Christopher {Heiter}}
\author[\Bochum, \Oxford, \RAPP]{{Mario R.} Hoerbe}
\author[\BUW]{Karl-Heinz {Kampert}}
\author[\Innsbruck, \Bochum, \RAPP]{Lukas {Merten}}
\author[\RWTH]{Gero {M\"{u}ller}}
\author[\Bochum, \Paris, \RAPP]{Patrick {Reichherzer}}
\author[\IKBU, \Moscow]{Andrey {Saveliev}}
\author[\Bochum, \RAPP]{Leander {Schlegel}}
\author[\Hamburg]{G\"unter {Sigl}}
\author[\DESY]{Arjen {van Vliet}}
\author[\VUB, \MPIFR]{Tobias {Winchen}}

\newcommand{\Radboud}{a}
\newcommand{\Bochum}{b}
\newcommand{\RAPP}{c}
\newcommand{\GSSI}{d}
\newcommand{\RWTH}{e}
\newcommand{\MPIFR}{f}
\newcommand{\Oxford}{g}
\newcommand{\BUW}{h}
\newcommand{\Innsbruck}{i}
\newcommand{\Paris}{j}
\newcommand{\IKBU}{k}
\newcommand{\Moscow}{l}
\newcommand{\Hamburg}{m}
\newcommand{\DESY}{n}
\newcommand{\VUB}{o}

\affiliation[\Radboud]{Radboud University Nijmegen, Department of Astrophysics/IMAPP, The Netherlands}
\affiliation[\Bochum]{Ruhr-Universität Bochum, Germany}
\affiliation[\RAPP]{Ruhr Astroparticle and Plasma Physics Center, Ruhr-Universität Bochum, Germany}
\affiliation[\GSSI]{Gran  Sasso  Science  Institute  (GSSI),  L’Aquila,  Italy}
\affiliation[\RWTH]{RWTH Aachen University, III. Physikalisches Institut A, Aachen, Germany}
\affiliation[\MPIFR]{Max Planck Institute for Radio Astronomy, Bonn, Germany}
\affiliation[\Oxford]{University of Oxford, Department of Astrophysics, Oxford, United Kingdom}
\affiliation[\BUW]{Bergische Universität Wuppertal, Department of Physics, Wuppertal, Germany}
\affiliation[\Innsbruck]{Institute for Astro- and Particle Physics, University of Innsbruck, Austria}
\affiliation[\Paris]{IRFU, CEA, Université Paris-Saclay, Gif-sur-Yvette, France}
\affiliation[\IKBU]{Immanuel Kant Baltic Federal University, Kaliningrad, Russia}
\affiliation[\Moscow]{Lomonosov Moscow State University,  Moscow, Russia}
\affiliation[\Hamburg]{II. Institute for Theoretical Physics, Hamburg  University, Hamburg, Germany}
\affiliation[\DESY]{DESY, Zeuthen, Germany}
\affiliation[\VUB]{Vrije Universiteit Brussel, Astrophysical Institute, Brussels, Belgium}

\emailAdd{crpropa@desy.de}

\abstract{
	The landscape of high- and ultra-high-energy astrophysics has changed in the last decade, in large part owing to the inflow of high-quality data collected by present cosmic-ray, gamma-ray, and neutrino observatories. At the dawn of the multimessenger era, the interpretation of these observations within a consistent framework is important to elucidate the open questions in this field. CRPropa 3.2 is a Monte Carlo code for simulating the propagation of high-energy particles in the Universe. This new version represents a step further towards a more complete simulation framework for multimessenger studies. Some of the new developments include: cosmic-ray acceleration, support for particle interactions within astrophysical sources, full Monte Carlo treatment of electromagnetic cascades, improved ensemble-averaged Galactic propagation, and a number of technical enhancements. Here we present some of these novel features and some applications to gamma- and cosmic-ray propagation.
	}

\FullConference{37$^{\rm{th}}$ International Cosmic Ray Conference (ICRC 2021)\\
		July 12th -- 23rd, 2021\\
		Online -- Berlin, Germany}

\begin{document}
\maketitle

\section{Introduction} \label{sec:intro}

In the last decade, a wealth of observations of energetic astrophysical phenomena were delivered with unprecedented precision. At ultra-high energies, the Pierre Auger and the Telescope Array collaborations reported observations of medium- and large-scale anisotropies in the arrival directions of ultra-high-energy cosmic rays (UHECRs) whose origins are yet unexplained~\cite{ta2014a, auger2017b, auger2018c, ta2020a, dimatteo2021a}. There are also noteworthy correlations between UHECRs and gamma-ray--emitting starburst galaxies~\cite{auger2018a}. The IceCube Neutrino Observatory detected a diffuse flux of neutrinos with $\sim$~PeV energies~\cite{icecube2013b}. Together with observatories operating from radio to gamma rays, IceCube also identified the blazar TXS~0506+056 as a potential source of PeV neutrinos~\cite{icecube2018b}. Observations of gamma rays with $\sim$~PeV energies hint at the existence of Galactic sources of high-energy CRs~\cite{hess2016a, tibet2019a, hawc2021a}.
Moreover, gravitational waves (GWs) became an integral part of the multimessenger paradigm in astrophysics, delivering pieces of information which combined with those obtained by other messengers have offered us glimpses into extreme events. 

Within this landscape, the interpretation of multimessenger observations is of paramount importance for fully grasping the phenomena at hand, and for answering the open questions in the field of cosmic rays~\cite{beckertjus2020a, alvesbatista2019d, kampert2014a}. This requires detailed theoretical knowledge of how each messenger is produced in astrophysical objects, how they can escape the environment wherein they are accelerated, how they propagate to Earth, and what are the multimessenger signatures associated to specific scenarios. For this purpose, we present here CRPropa~3.2.

CRPropa~3.2 extends the applicability of the code to a variety of applications beyond UHECR astrophysics, which range from the propagation of CRs in the Galaxy to gamma-ray and neutrino astronomy. Here we introduce the capabilities of CRPropa in section~\ref{sec:crpropa}, followed by a description of some of the improvements made in this new version in section~\ref{sec:newFeatures}. We illustrate some applications of the code in section~\ref{sec:applications}, and briefly address the current status of CRPropa~3.2 and future developments in section~\ref{sec:outlook}.

\section{The CRPropa Framework} \label{sec:crpropa}

CRPropa~3~\cite{alvesbatista2016a, merten2017a} is a flexible framework for high-energy astrophysical studies. It is written in C++ with Python bindings. The code structure is fully modular, handling each ingredient of the transport problem separately through the usage of individual modules that describe, e.g., sources, observer, particles, interactions, and magnetic fields. Each module acts on properties of an object from the \codeStyle{Candidate} class, which stores all the relevant information about a particle such as its position, energy, nature (type of particle), and momentum. At each step of propagation the modules update these properties according to prescriptions which may involve, for instance, kinematic changes due to a given physical process such as interactions, magnetic deflections, etc. Particles are tracked until some specified breaking condition is met, which may include trajectory length constraints or detection, for example.

CRPropa simulations can be processed in parallel, with shared memory, via OpenMP. This enables fast computations covering large parameter spaces. Moreover, depending on the specific problem at hand, one can make use of the fact that CRPropa is a Monte Carlo code and simply apply weights to a single set of simulations to create a family of scenarios. This greatly optimises performance and makes it possible to probe an even larger parameter space to compare different scenarios of interest.

\section{New Features} \label{sec:newFeatures}

In CRPropa~3.2 we have extended the physics capabilities of the code by introducing a number of new modules to treat phenomena that so far could not be modelled with the code. Below we list some of them.

\medskip
\paragraph{Acceleration.} CRPropa~3.2 contains modules to simulate the acceleration of charged particles scattering off magnetic-field irregularities. In particular, first- and second-order Fermi acceleration are available, generalising the implementation of ref.~\cite{winchen2018a}. 

\medskip
\paragraph{New Photon Production Channels.} Processes such as photodisintegration of nuclei and radiative decays can produce (ultra-)high-energy photons. For this reason, we added such production channels to CRPropa, together with elastic scattering of CR nuclei off background photons~\cite{heiter2018a}. The latter, in particular, does not incur substantial energy losses for cosmic rays, but can be important to obtain a consistent multimessenger picture. 

\medskip
\paragraph{Electromagnetic Cascades.} So far, the propagation of photons in CRPropa was done using \emph{external} packages such as the transport-equation solver DINT~\cite{lee1998a} or the Monte Carlo code EleCa~\cite{settimo2015a}. For consistency and to enable a full three-dimensional treatment of photon propagation, we implemented all relevant electromagnetic processes for high-energy photons and electrons natively within CRPropa. These interactions can occur many times during the cascade process, leading to a rapid growth in the number of particles in the simulation, which renders a Monte Carlo treatment over a wide range of energies impractical. To enable faster simulations, we also implemented a thinning procedure following refs.~\cite{kachelriess2012a, alvesbatista2016b}, leading to speedups of up to four orders of magnitude. 

\medskip
\paragraph{Ensemble-Averaged Propagation.} The propagation of charged particles can be diffusive, making the particle-by-particle treatment of propagation computationally expensive. One example is the propagation of CRs with energies $E \lesssim 10 \; \text{PeV}$ in the Milky Way. Since CRPropa~3.1~\cite{merten2017a}, an ensemble treatment of this problem has been part of CRPropa by solving stochastic differential equations (SDEs) corresponding to the Parker transport equation. These SDEs describe the time evolution of pseudo-particles representing phase-space elements. This framework was extended to include macroscopic effects due to bulk motion such as advection and adiabatic compression/expansion. 

\medskip
\paragraph{Magnetic Field Models.} We implemented new generic magnetic-field models, namely a single-mode helical~\cite{alvesbatista2019b} and a helical turbulent~\cite{alvesbatista2016b} field. New models for the Galactic magnetic field (GMF) were added, in addition to the already existing ones. This includes modifications of the model by Jansson \& Farrar~\cite{jansson2012a, jansson2012b} such as those from refs.~\cite{planck2016c, kleimann2019a}, as well other models for the halo field~\cite{jokipii1987a} and the central Galactic zone~\cite{guenduez2020a}.

\bigskip 
In addition to the new physics features, several technical improvements were made aiming at optimising performance and accuracy.

\medskip
\paragraph{Magnetic Field Improvements.} We have redesigned the way in which CRPropa handles stochastic magnetic fields. The discrete three-dimensional turbulent fields were extended to allow for more sophisticated magnetic power spectra, now including a smooth bend-over of the spectrum and different indices for the injection and cascade regions. This grid-based method is useful and can be applied to a wide range of problems, but an alternative approach that computes magnetic fields on the fly can be more suitable for other problems. A notable example is the propagation of charged particles when their Larmor radii are smaller than the coherence length of the magnetic field (see, e.g., refs.~\cite{reichherzer2020a, dundovic2020a, reichherzer2021a}). This is implemented in CRPropa~3.2 using plane waves, following ref.~\cite{tautz2013a}. 


\medskip
\paragraph{Targeting.} One of the main difficulties in three- and four-dimensional Monte Carlo simulations of particle propagation over large distances is the detection efficiency of the collection surfaces representing the detector. Particles emitted by a given source are tracked, deflected by intervening magnetic fields, but often they do not reach a small target that represents the observer within this set-up, making this type of simulation computationally costly. CRPropa~3.2 features a novel algorithm which updates the direction of the particles emitted by a source according to the (non-)detections of smaller sample of events. This can lead to substantial performance gains depending on the scenario of interest, of up to four orders of magnitude~\cite{jasche2019a}.

\medskip
\paragraph{Boris Push Algorithm.} The transport of individual charged particles across a magnetised environment was currently done in CRPropa using the Cash-Karp algorithm~\cite{cash1990a}. In CRPropa~3.2 we implemented an additional algorithm, the Boris push~\cite{boris1972a}, widely used in plasma physics thanks to its speed and accuracy. Both algorithms support adaptive step sizes.

\medskip
\paragraph{Enhanced Interpolation Methods.} Turbulent magnetic fields are often stored in grids before usage. When modelling the propagation of charged particles, the magnetic field is interpolated and the trajectory of the particle is corrected according to this value. The grids have to cover large distances whilst resolving small scales, such that poor interpolation methods could lead to miscalculation of particle trajectories~\cite{schlegel2020a}. For this reason, three-dimensional grids in CRPropa can now be interpolated using more alternative methods such as nearest-neighbour, trilinear, and tricubic interpolations. The optimal choice is a compromise between accuracy and performance, and it depends on the specific problem.

\section{Applications} \label{sec:applications}

In this section we present a few applications of CRPropa to various problems. Some of the examples highlight the novel features of CRPropa~3.2.

\subsection{Cosmic-Ray Diffusion at the Edge of the Galaxy} \label{ssec:ex_galactic}

\begin{figure}[htb]
	\centering
	\includegraphics[width=0.75\textwidth]{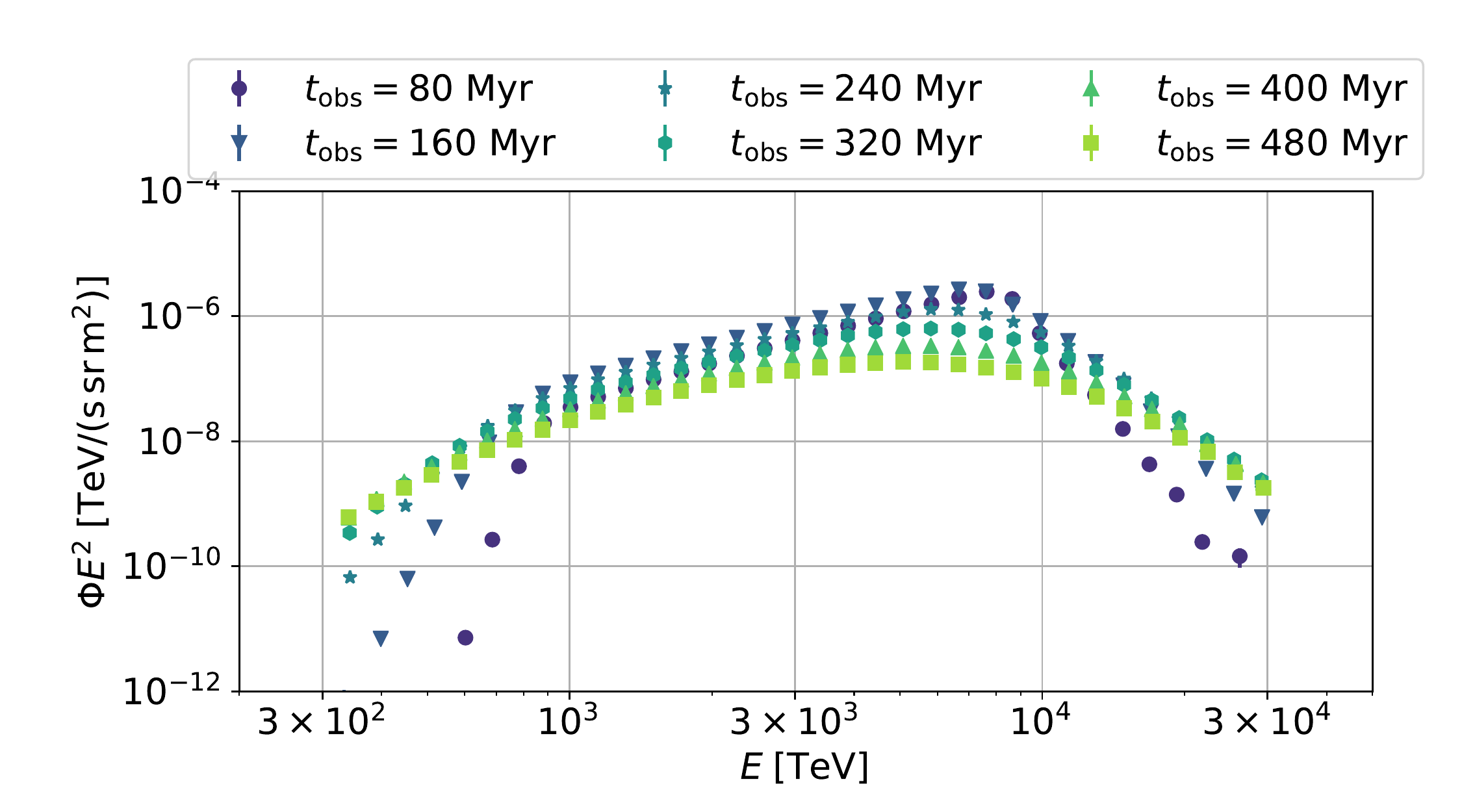}
	\caption{Energy spectrum $\Phi$ of CRs from the Galactic wind termination shock at an observer with radius 10~kpc around the Galactic center assuming a perpendicular-to-parallel diffusion ratio $\epsilon=0.1$ and Kraichnan turbulence spectrum. For details see ref.~\cite{merten2018a}.}
	\label{fig:galactic}
\end{figure}

CRPropa can be used to study the propagation of CRs in the Milky Way and its outskirts including processes such as advection and adiabatic effects. Here, we show an example in which a Galactic termination shock re-accelerates over $10^8$~years 1--10~PeV CRs blown out by the Galactic wind. Naturally, the energy gain depends on the magnetic field of the Galactic halo which. In this case, it is the recently-implemented Archimedean spiral. These results have immediate implications for understanding the end of the Galactic CR spectrum and the transition to the extra-galactic component. Figure \ref{fig:galactic} shows the time evolution of the CR spectrum just outside the Galaxy. For more details see ref.~\cite{merten2018a}.

\subsection{Cosmological Propagation of Gamma Rays} \label{ssec:ex_gamma}

\begin{figure}[!htb]
	\centering
	\includegraphics[width=0.495\textwidth]{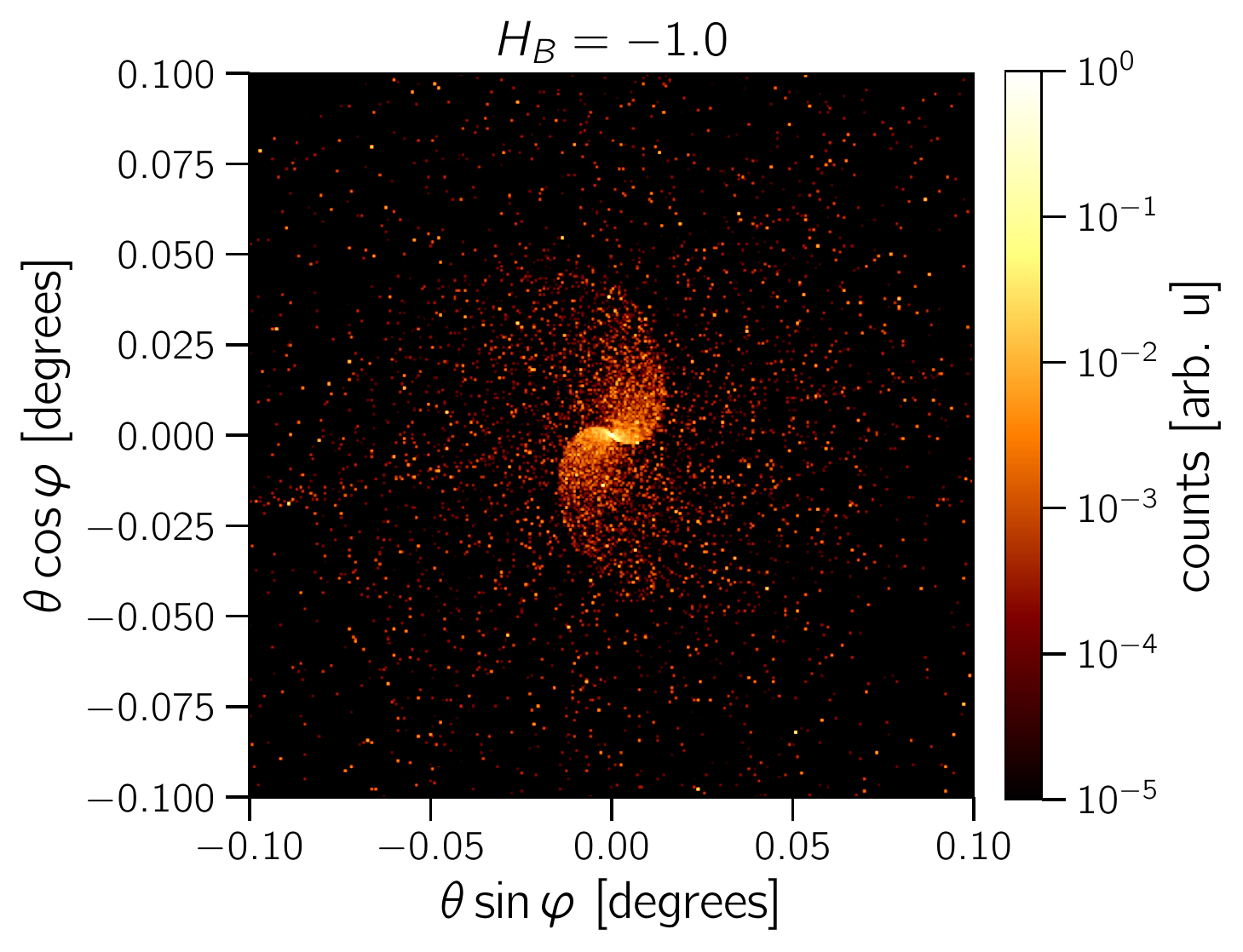}
	\includegraphics[width=0.495\textwidth]{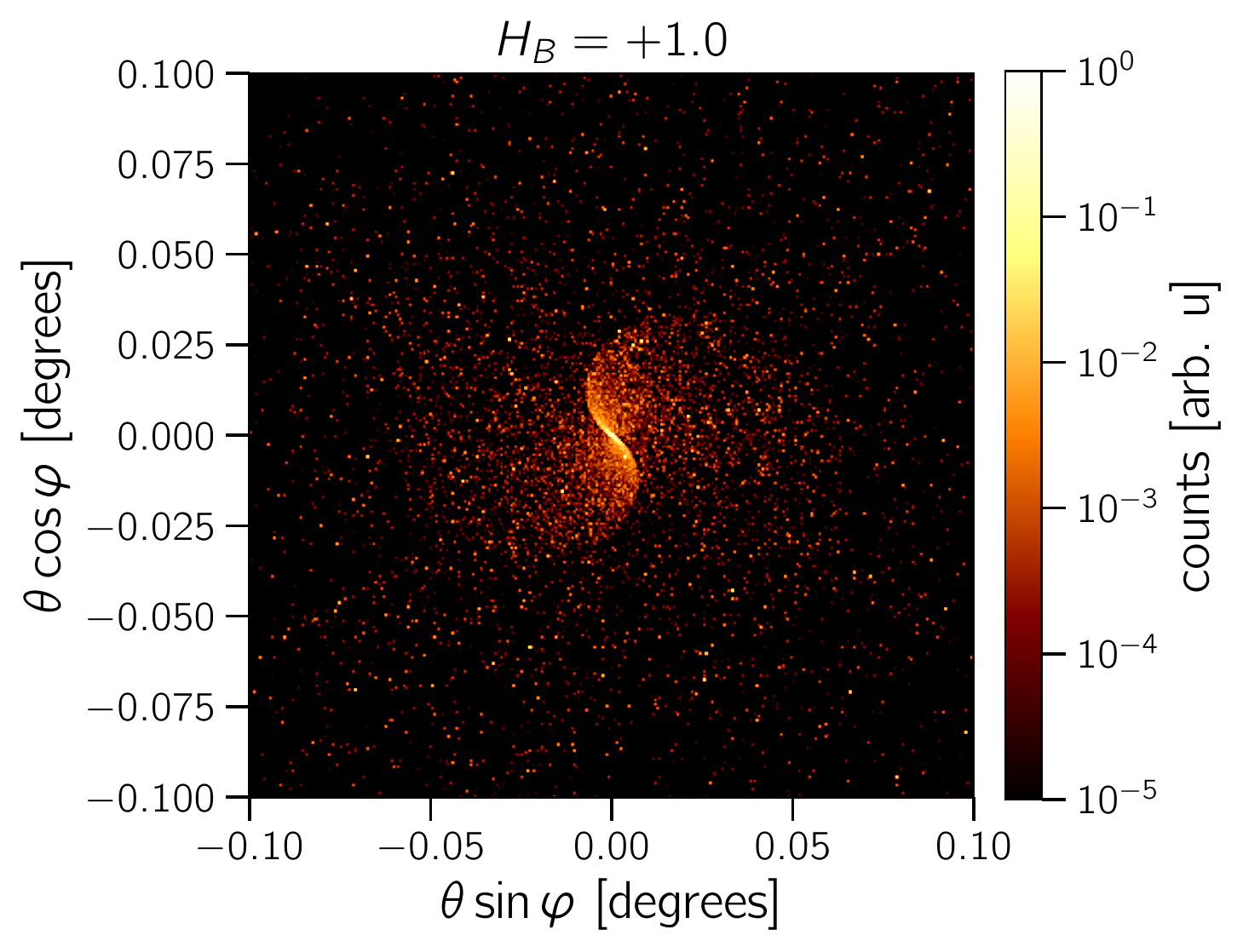}
	\caption{The figures shows arrival directions of gamma rays emitted from a hypothetical source located at $z=0.08$ as observed at Earth. The emission spectrum is a power-law of the form $E^{-1.5}$ with an expontial cut-off at $E_\text{max} = 100 \; \text{TeV}$. The magnetic-field strength is $B=10^{-15} \; \text{G}$ with coherence length $L_B = 200 \; \text{Mpc}$, assuming a Batchelor spectrum. The panels correspond to a maximally negative ($H_B = -1$) and a maximally positive ($H_B = 1$) helicity of the IGMFs, respectively. For more details see ref.~\cite{alvesbatista2021a}.}
	\label{fig:gamma}
\end{figure}

With the new modules for electromagnetic interactions recently added to CRPropa, it is possible to simulate a wide variety of scenarios. An interesting application is the study of gamma-ray induced electromagnetic cascades in intergalactic space. This includes effects such as absorption by the extragalactic background light (EBL) and the deflection of electron-positron pairs by intergalactic magnetic fields (IGMF)~\cite{alvesbatista2021a}. One such example is shown in fig.~\ref{fig:gamma}, in which we simulate the propagation of gamma rays through turbulent helical magnetic fields --- another novel feature of CRPropa~3.2. 

\subsection{Multimessenger studies} \label{ssec:ex_multimessenger}

\begin{figure}[!ht]
	\centering
	\includegraphics[width=0.495\textwidth]{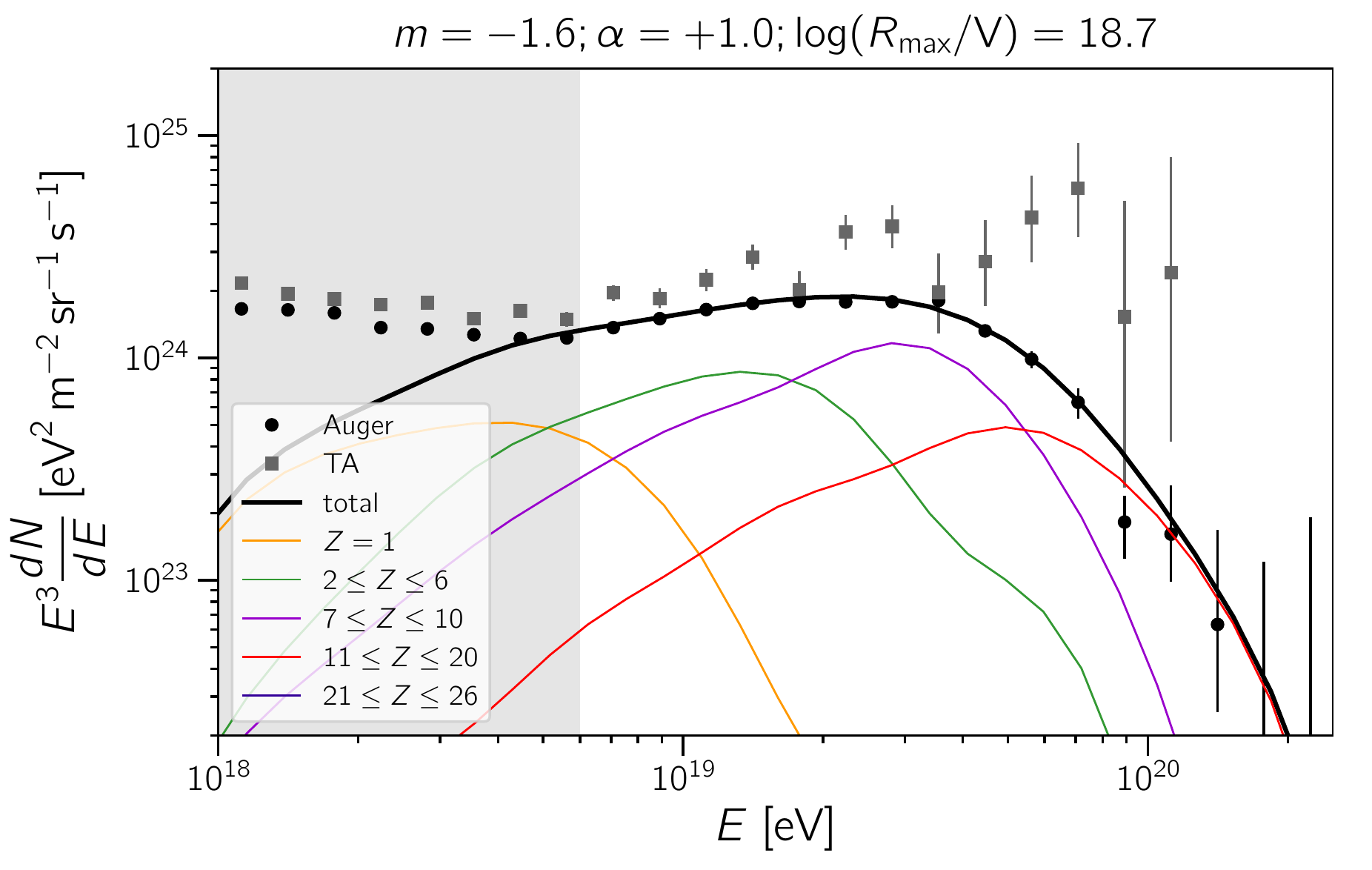}
	\includegraphics[width=0.495\textwidth]{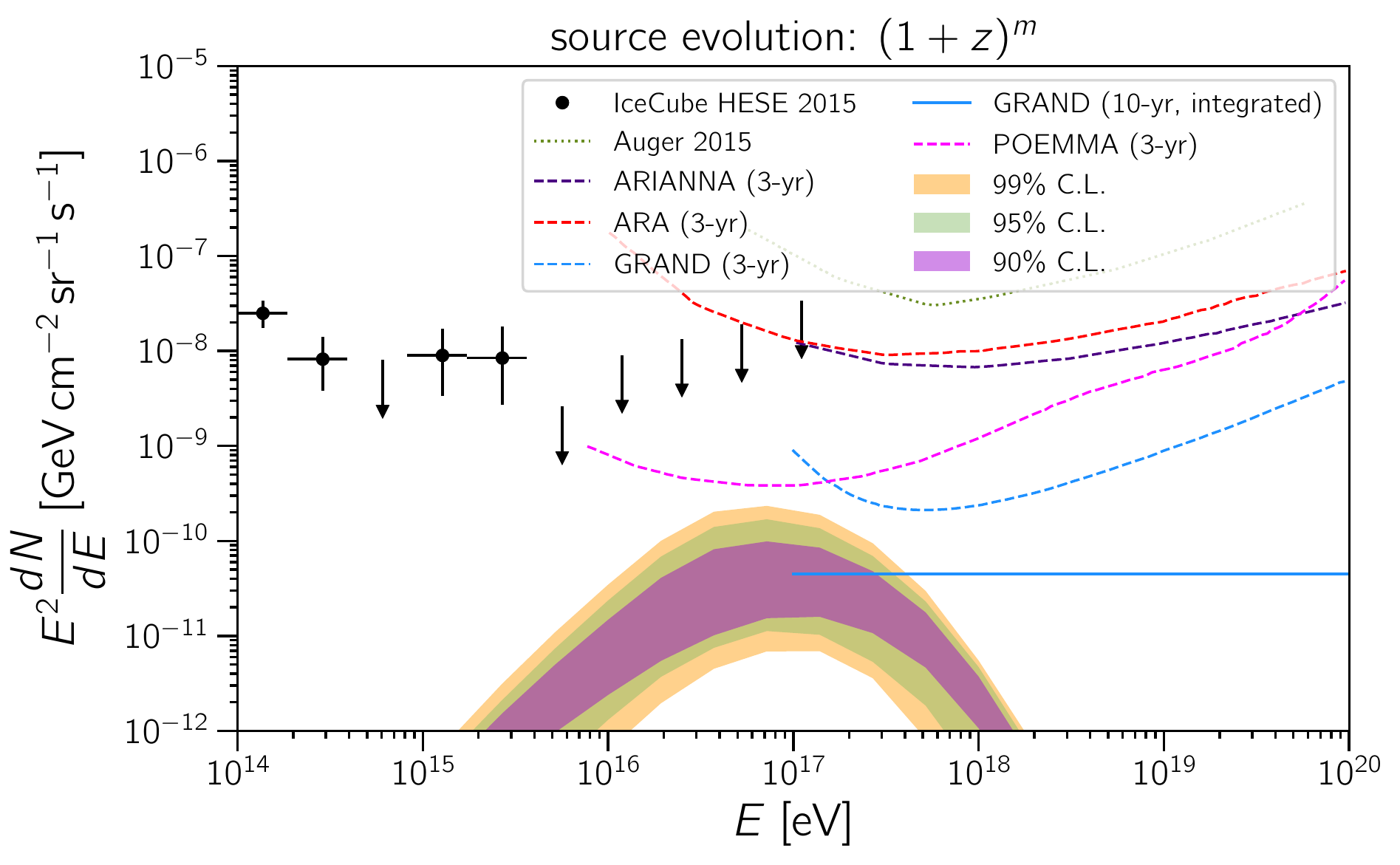}
	\caption{\emph{Left:} Fit to the spectrum measured by Auger assuming a power-law spectrum with spectral index $\alpha$ and an exponential cut-off at $R_\text{max}$ for sources evolving as $(1 + z)^m$. \emph{Right:} Estimated cosmogenic neutrino fluxes obtained from the fit for different confidence levels. Figures taken from ref.~\cite{alvesbatista2019a}.}
	\label{fig:multimessenger}
\end{figure}

CRPropa can self-consistently compute fluxes of multiple messengers compatible with UHECR observations, as done in, e.g., refs.~\cite{auger2017a, alvesbatista2019a}. In the left panel of fig.~\ref{fig:multimessenger} we illustrate these capabilities by presenting a fit to the observations by Auger at energies $E \gtrsim 10^{18.7} \; \text{eV}$. We consider all relevant photohadronic interactions, namely photopion production, photodisintegration and Bethe-Heitler pair production with the CMB and the EBL. The cosmogenic neutrino fluxes associated with this scenario are shown in the right panel. For more details, the reader is referred to ref.~\cite{alvesbatista2019a}. Besides fluxes, correlations of arrival directions between multiple messengers can be computed self-consistently as well, as done in, e.g., ref.~\cite{vanvliet2019a}.

\subsection{Particle Acceleration} \label{ssec:ex_acceleration}

\begin{figure}[h!]
    \centering
	\includegraphics[width=0.7\textwidth]{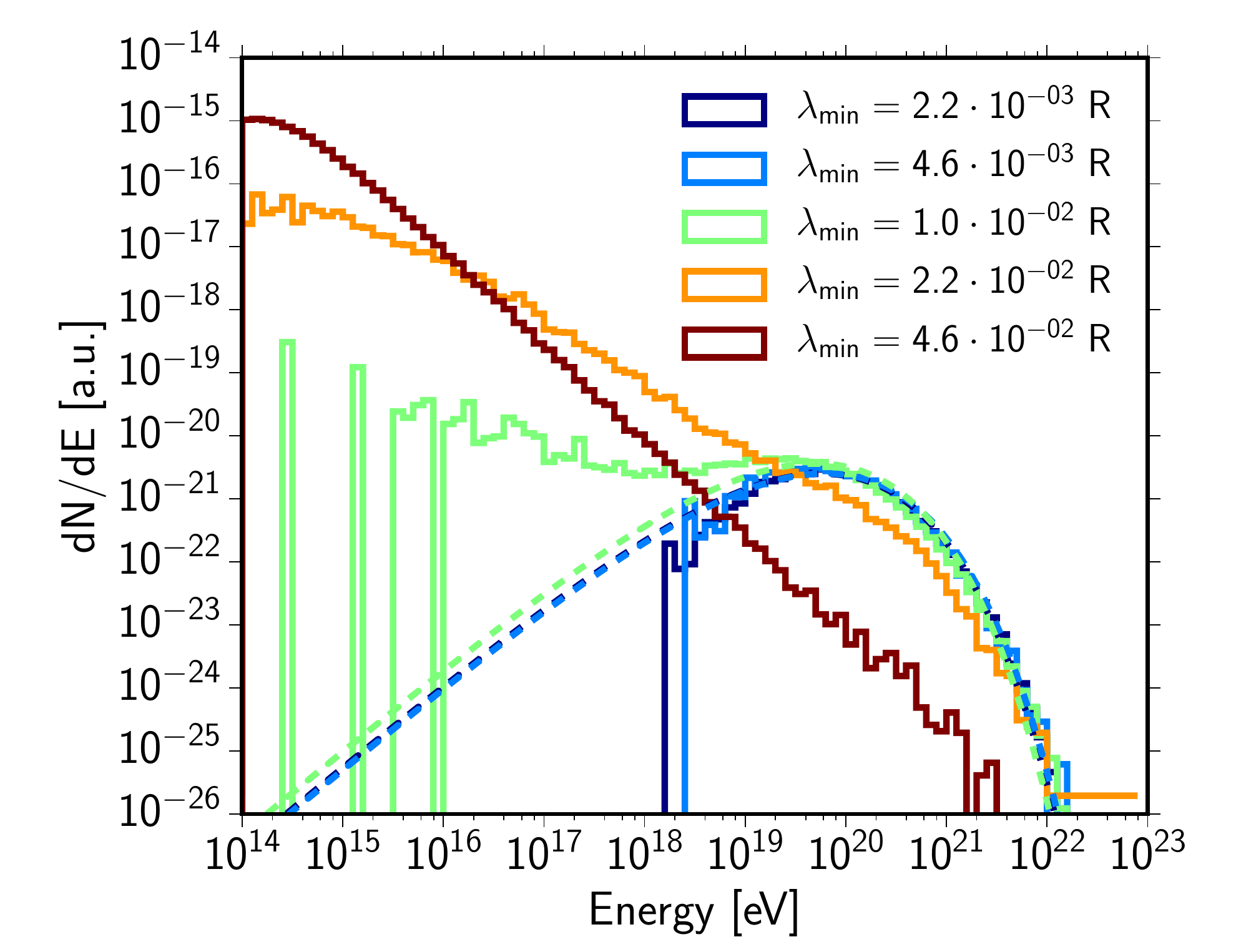}
	\caption{Spectrum of protons accelerated via the second-order Fermi mechanism. Here particles are injected at the centre of a sphere of radius $R$. $\lambda_\text{min}$ denotes the step length used in the simulation. The dashed lines show the theoretical predictions. Figure taken from ref.~\cite{winchen2018a}.}
	\label{fig:acceleration}
\end{figure}

To illustrate how CRPropa can treat particle acceleration, we consider the case of CRs being accelerated by a source with a Kolmogorov magnetic field power spectrum. The velocity of the scattering centres is assumed to be $\beta = 0.1$. The simulation takes as argument the scattering length by plasma waves ($\lambda_\text{min}$), which is approximately constant for Larmor radii much smaller than the maximum magnetic turbulence injection scale. The spectrum of accelerated CRs is shown in fig.~\ref{fig:acceleration} for several values of $\lambda_\text{min}$. 

\section{Conclusions and Outlook} \label{sec:outlook}

To interpret current astrophysics observations at high- and ultra-high energies, detailed theoretical models with multiple types of messengers are necessary to obtain a self-consistent picture of the Universe across many decades in energy. CRPropa~3.2 is a simulation framework with a broad scope that enables such studies. Our recent developments make substantial technical improvements to optimise the code performance in simulations with magnetic fields, whether in the Milky Way or outside of it. They also extend the range of possible applications of CRPropa by treating particle acceleration, by implementing powerful tools for transporting cosmic rays in the diffusion regime, by adding new channels for the production of photons, and by implementing a native treatment of electromagnetic interactions. 

Here, we illustrated some applications of CRPropa to a variety of problems involving the propagation of CRs, gamma rays, electrons, and neutrinos with energies from $\sim$~GeV (or TeV, in the case of CRs) up to ZeV. This consolidates CRPropa as a comprehensive framework for high-energy studies in various astrophysical environments. 

\bigskip
Some of the latest developments were already released in the current version, whereas the others will be part of the upcoming CRPropa~3.2. More information about CRPropa can be found in our webpage: 
\url{https://crpropa.desy.de}.

\section*{Acknowledgements}

RAB acknowledges the support from the Radboud Excellence Initiative. LM acknowledges financial support from the Austrian Science Fund (FWF) under grant agreement number I~4144-N27.
KHK, LS and JT acknowledge support from the Deutsche Forschungsgemeinschaft for the project \textit{ Multi-messenger probe of Cosmic Ray Origins (MICRO)}, grant numbers TJ\,62/8-1 and KA\,710/5-1. The work of AS is supported by the Russian Science Foundation under grant no.~19-11-00032.
AvV acknowledges support from the European Research Council (ERC) under the European Union’s Horizon 2020 research and innovation program (Grant No.~646623).


\end{document}